\documentclass[letter]{IEEEtran}
\IEEEoverridecommandlockouts
\usepackage{cite}
\usepackage{amsmath,amssymb,amsfonts}
\usepackage{algorithm}
\usepackage{algpseudocode}
\usepackage{graphicx}
\graphicspath{{./images/}}
\DeclareGraphicsExtensions{.pdf,.jpeg,.png,.eps}
\usepackage[caption=false,font=normalsize,labelfont=sf,textfont=sf]{subfig}
\usepackage{textcomp}
\usepackage{xcolor}
\newcommand{\tildebf}[1]{\Tilde{\mathbf{#1}}}
\def\BibTeX{{\rm B\kern-.05em{\sc i\kern-.025em b}\kern-.08em
    T\kern-.1667em\lower.7ex\hbox{E}\kern-.125emX}}
    
\begin{document}

\title{Resource Allocation for Mixed Numerology NOMA\\
}

\author{\IEEEauthorblockN{Stephen McWade,
Mark F. Flanagan,
Juquan Mao,
Lei Zhang and 
Arman Farhang}
\thanks{S. McWade and M. Flanagan are with the School of Electrical \& Electronic Engineering, University College Dublin, Ireland (email: stephen.mcwade@ucdconnect.ie and mark.flanagan@ieee.org). J. Mao is with the Institute for Communications Systems, University of Surrey, UK (email: juquan.mao@surrey.ac.uk). L. Zhang is with the School of Engineering, University of Glasgow, UK (email: Lei.Zhang@glasgow.ac.uk). A. Farhang is with the Department of Electronic Engineering, National University of Ireland Maynooth, Ireland (email: Arman.Farhang@mu.ie).}}

\maketitle

\begin{abstract}
6G wireless networks will require the flexibility to accommodate an extremely diverse set of service types. This necessitates the use of mixed numerologies to accommodate different quality of service (QoS) requirements. Non-orthogonal multiple access (NOMA) techniques can potentially be used to accommodate users with different numerologies while also gaining the performance benefits associated with NOMA. To achieve the full performance benefits of a mixed numerology NOMA (MN-NOMA) system, resource allocation among the users is paramount.  However, the coexistence of mixed numerologies changes the nature of the interference that each user experiences. This means that techniques used in single-numerology NOMA (SN-NOMA) are no longer sufficient. In light of this, we approach the problem of optimizing subcarrier and power allocation for maximizing the spectral efficiency of MN-NOMA while considering a minimum rate constraint for each user. In this letter, we propose a two-stage sub-optimal approach to solve the problem. We present numerical results which show the superiority of our proposed method over existing benchmark schemes in both spectral efficiency and fairness.
\end{abstract}


\section{Introduction}

Future mobile networks will require high flexibility and the ability to simultaneously provide service to multiple users with different service types and quality of service (QoS) requirements (e.g. ultra-reliable and low-latency communications (URLLC), massive machine-type communications (mMTC), vehicle-to-everything (V2X) communications, etc.). The sixth generation of communication networks (6G) are envisioned to have an even wider variety of service types\cite{Saad6G_2020} and will require an even higher degree of flexibility to achieve this. Waveform numerology, which here refers to the waveform design parameters, becomes extremely important for achieving the diverse multi-service vision for 6G. For example, mMTC may require a narrow subcarrier spacing to support delay-tolerant devices, whereas V2X will require a wider subcarrier spacing or shorter symbol duration to improve robustness against Doppler spread \cite{5G_Slicing}. A one-size-fits all numerology is obviously very difficult (if not impossible) to design. This then leads to the problem of how best to accommodate services with mixed numerologies and achieve the required flexibility. 

The standard approach to this problem is to separate the system bandwidth into smaller adjacent bandwidth parts (BWPs) with each BWP having a different numerology for its service type. Throughout this letter, we refer to this approach as mixed numerology orthogonal multiple access (MN-OMA). However, the OFDM subcarriers of mixed numerologies are not orthogonal to each other and this causes inter-numerology interference (INI) which in turn degrades system performance \cite{zhang_2018}. An alternative to this approach is to use mixed numerology non-orthogonal multiple access (MN-NOMA). In this scenario, users with different numerologies share time and frequency resources and are multiplexed in another domain such as the power or the code domain \cite{DaiSurveyNOMA}. Superposition coding and successive interference cancellation (SIC) are used to accurately decode the user signals.  Our previous work \cite{McWade2020} showed that MN-NOMA can accommodate users with different numerologies while providing improved spectral efficiency (SE) over the MN-OMA approach.

The topic of resource allocation in NOMA has received significant attention in recent years. The authors of  \cite{AlImariUpNOMA2015} and \cite{FuNOMA2018} both address the problem of power allocation for maximizing efficiency in NOMA. Resource allocation is equally important for mixed numerology NOMA in order to achieve the full potential of the system. However, the presence of mixed numerologies changes the nature of the interference experienced by the users.  This means that assumptions used in solutions for single numerology NOMA (SN-NOMA), such as \cite{ZengUplinkNoma}, are not applicable in MN-NOMA. 

While there has been research on resource allocation for MN-OMA \cite{Mao2020}, to the best of our knowledge there has been little research on the topic of optimizing resource allocation for MN-NOMA. The authors of \cite{Abusabah2018NOMAFM} tackle the problem of power allocation of a multinumerology NOMA system with a constraint on system fairness. However, \cite{Abusabah2018NOMAFM} uses an exhaustive search method to solve the optimization problem, which is clearly impractical for real-world applications as it scales very poorly with the numbers of users. The authors of \cite{Wang_MN_NOMA2020} consider the allocation of time-frequency resource blocks for MN-NOMA but do not consider power allocation which is an important topic for an MN-NOMA system. Another gap in the existing works on both MN-OMA and MN-NOMA is that they almost exclusively consider a small number of users (e.g., 2 or 3 users only\cite{McWade2020,Mao2020,Abusabah2018NOMAFM}).

 This letter addresses the above mentioned gaps in the literature with the following contributions:

\begin{itemize}
    \item We outline a generic analytical model for mixed numerology uplink NOMA where any number of users with different numerologies transmit over multipath fading channels and are decoded via SIC at the receiver.
    
    \item We formulate an optimization problem for maximizing the spectral efficiency of this MN-NOMA system subject to a minimum rate requirement at each user. Additionally, in order to reduce SIC complexity and error propagation effects, we consider a scenario where a limitation is placed on the number of users that can occupy a single subcarrier.
    
    \item Since the optimization problem is NP-hard, we utilize a two-stage sub-optimal approach to solve it. Stage 1 uses an iterative greedy algorithm to allocate subcarriers to users for a fixed initial power allocation. Stage 2 uses an successive convex approximation (SCA) based approach to optimize the power allocation for MN-NOMA users. 
    
\end{itemize}

\subsubsection*{Notations} Superscripts ${(\cdot)^{\rm{T}}}$ and ${(\cdot)^{\rm{H}}}$ denote transpose and Hermitian transpose, respectively. Bold lower-case characters are used to denote vectors and bold upper-case characters are used to denote matrices. $\mathbf{X}=\rm{diag}(\mathbf{x})$ is a diagonal matrix with the elements of the vector $\mathbf{x}$ on its main diagonal. $\mathbf{x}=diag(\mathbf{X})$ is a column vector whose elements consist of the main diagonal of the matrix $\mathbf{X}$, and $\otimes$ represents Kronecker product. The $p\times{p}$ identity matrix and $p \times q$ all-zero matrix are  denoted by $\mathbf{I}_p$ and $\mathbf{0}_{p\times{q}}$, respectively.

\section{System Model}

We consider a mixed numerology multi-carrier uplink NOMA system with $K$ users. Each user uses cyclic prefix orthogonal frequency division multiplexing (CP-OFDM)  modulation each with their own specified subcarrier spacing $\Delta f_i$, $i\in\{1,\ldots,K\}$, i.e., their own numerology. The subcarrier spacings of users $i$ and $j$ are related as
$\Delta_{i,j}=\frac{\Delta{}f_{i}}{\Delta{}f_{j}} = \frac{q_i}{q_j},$
where \(q_i = 2^{\mu_i}\), \(\mu_i \in \{0,1,2,3...\}\), is the scaling factor of user $i$'s numerology as per 5G NR \cite{3gpp_TS38211}. 
User $i$ has $N_{i} = \frac{B}{\Delta f_i}$ subcarriers available, where $B$ is the system bandwidth. Each user has a corresponding CP length $N_{\mathrm{cp},i}$, which is scaled depending on the user numerology to maintain alignment of the time domain symbols. The total symbol length for user $i$ is therefore $N_{\mathrm{T},i} = N_i + N_{\mathrm{cp},i}$ . 

We define the power allocation vector of user $i$ as ${\mathbf{p}_{i}} = \left[\sqrt{{p}_{i,0}}\ \sqrt{{p}_{i,1}}\ \dots\ \sqrt{{p}_{i,N_{i}-1}} \right]^{\rm{T}}$ where ${p}_{i,n}$ is the power allocated to subcarrier $n \in \{0,1,...N_{i}-1\}$ of user $i$.
We also define a vector $\mathbf{p}=\left[\mathbf{p}_{1}^{\rm{T}}\  \mathbf{p}_{2}^{\rm{T}}\ \dots\ \mathbf{p}_{K}^{\rm{T}} \right]^{\rm{T}}$ as the power allocation vector  for the whole system. In order to accommodate a limit on the number of users that share a subcarrier, we denote the $N_i \times 1$ subcarrier allocation vector of user $i$ as $\mathbf{x}_i = \left[{x}_{i,0}\ {x}_{i,0}\ \dots\ {x}_{i,N_{i}-1} \right]^{\rm{T}}$. The element $x_{i,n} = 1$ if subcarrier $n$ is allocated to user $i$ and $x_{i,n} = 0$ if it is not. We define a vector $\mathbf{X}=\left[\mathbf{x}_{1}^{\rm{T}}\  \mathbf{x}_{2}^{\rm{T}}\ \dots\ \mathbf{x}_{K}^{\rm{T}} \right]^{\rm{T}}$ as the subcarrier allocation vector for the whole system.

The transmitted symbol for user $i$ is given by
\begin{equation}
    \mathbf{s}_{i} = {\mathbf{A}}_{\mathrm{cp},i}{\mathbf{F}_{i}^{\rm{H}}}\mathrm{diag}(\mathbf{x}_i)\mathrm{diag}(\mathbf{p}_i){\mathbf{d}_i},
    \label{eq:1}
\end{equation}
where $\mathbf{F}_{i}$ is the $N_i$-point unitary discrete Fourier transform (DFT) matrix of user $i$ with elements $F_i[l,k]=\frac{1}{\sqrt{N_i}}e^{-j\frac{2\pi}{N_i}lk}$. The matrix ${\mathbf{A}}_{\mathrm{cp},i} = \left[\mathbf{I}_{\rm{cp},i}, \mathbf{I}_{N_{i}}\right]^{\rm{T}} $ is the CP addition matrices of user $i$, where $\mathbf{I}_{\rm{cp},i}$ is composed of the final $N_{\mathrm{cp},i}$ columns of $\mathbf{I}_{N_{i}}$. The vector ${\mathbf{d}_i}$ is the vector of data-bearing symbols for user $i$. For each user's channel, we consider a linear time invariant channel model with channel impulse response $\bar{\mathbf{h}}_i = [\bar{h}_{i,0}, \ldots, \bar{h}_{i,L_i-1}]^{\rm T}$ where $L_i$ is the channel length. We define $\mathbf{H}_{i}$ as the Toeplitz channel matrix with first column equal to $\left[(\bar{\mathbf{h}}_i)^{\rm{T}}, \mathbf{0}_{1\times(N_{T,i}-L^{(i)}-1)}\right]^{\rm{T}}$ and first row $\left[\bar{h}_{i,0}, \mathbf{0}_{(1)\times(N_{T,i}-1)}\right]$. Each user's signal passes through its respective channel and these signals are superimposed at the base station. The user signals are then decoded at the base station using SIC. Without loss of generality, we assume that users are decoded in the order of their indices, i.e., user 1 is decoded first, then user 2, and so on until user $K$ is decoded last. When a user is decoded, it only experiences interference from the users yet to be decoded. Thus, the presence of mixed numerologies changes the interference experienced by the users. 

We consider a 2-user case as an example, where user $i$ experiences interference from user $j$. If $\Delta{}f_{i}<\Delta{}f_{j}$ then the time domain symbol of user $i$ overlaps with $\Delta_{j,i}$  time domain symbols of the user $j$. The received signal is given by 
\begin{equation}
    \mathbf{r}= \mathbf{H}_{i}\mathbf{s}_{i} + \mathbf{H}_{i}\tildebf{s}_j + \mathbf{w}
    \label{eq:2}
\end{equation}
where $\mathbf{w} \sim \mathcal{CN}(\mathbf{0}, {\sigma}^2\mathbf{I})$ represents additive white Gaussian noise. The vector $\tildebf{s}_j$ is the concatenation of the $\Delta_{j,i}$ overlapping user $j$ symbols. This concatenated symbol is given by
\begin{equation}
    \tildebf{s}_j = \left[\mathbf{I}_{\Delta_{j,i}}\otimes({\mathbf{A}}_{\mathrm{cp},j}{{\mathbf{F}}_{j}^{\rm{H}}}\mathrm{diag}(\mathbf{x}_j)\mathrm{diag}(\mathbf{p}_j))\right]\tildebf{d}_j
    \label{eq:3}
\end{equation}
where $\tildebf{d}_j$ is a vector of concatenated data-bearing symbols of user $j$. The decoded user $i$ signal is given by
\begin{equation}
    \mathbf{y}_i = \mathbf{F}_{i}\mathbf{R}_{\mathrm{cp},i}\mathbf{r}
    \label{eq:4}
\end{equation}
where $\mathbf{R}_{\mathrm{cp},i} = \left[\mathbf{0}_{N_i\times{N_{\mathrm{cp},i}}}, I_{N_i}  \right]$ is the CP removal matrix of user $i$. Using (\ref{eq:1}) and (\ref{eq:2}), (\ref{eq:4}) can be expanded as
\begin{equation}
    \mathbf{y}_i =\mathbf{\Psi}_i\mathrm{diag}(\mathbf{x}_i)\mathrm{diag}(\mathbf{p}_i){\mathbf{d}_i}+ \mathbf{F}_{i}\mathbf{R}_{\mathrm{cp},i}\mathbf{H}_{i}\tildebf{s}_j + \boldsymbol{\omega}_i,
    \label{eq:5}
\end{equation}
where $\mathbf{\Psi}_i$ is a square diagonal matrix with diagonal elements equal to the frequency response of the channel for user $i$, i.e., $\mathbf{F}_{i}\mathbf{R}_{\mathrm{cp},i}\mathbf{H}_{i}{\mathbf{A}}_{\mathrm{cp},i}{\mathbf{F}_{i}^{\rm{H}}}=\mathrm{diag}(\mathbf{h}_i)$ where $\mathbf{h}_i=\mathbf{F}_{i}[\bar{\mathbf{h}}_i,\mathbf{0}_{(N_i-L_i)\times 1}]^{\rm T}$. An interference matrix can be calculated from the second term of (\ref{eq:5}) as
\begin{equation}
        \boldsymbol{\Gamma}^{(i \leftarrow j)} = \mathbf{F}_{i}\mathbf{R}_{i}\mathbf{H}_{j}\left[\mathbf{I}_{\Delta_{j,i}}\otimes({\mathbf{A}}_{j}{{\mathbf{F}}_{j}^{\rm{H}}}\mathrm{diag}(\mathbf{x}_j)\mathrm{diag}(\mathbf{p}_j))\right].
\end{equation}
\(\boldsymbol{\Gamma}^{(i \leftarrow j)}\) is an $N_i\times{\Delta_{j,i}}N_j$ matrix where the $(n,o_m)$-th element \({\Gamma}_{n,o_m}^{(1 \leftarrow 2)}\) contains the INI weight on subcarrier $n$ of user $i$ from the the $o$-th subcarrier of the $m$-th overlapping symbol of user $j$ where $o_m = m(N_j)+o$ for $m=0,\ldots,\Delta_{i,j}-1$ and $o=0,\ldots,N_j-1$.

Alternatively, if $\Delta{}f_{i}>\Delta{}f_{j}$ then there are $\Delta_{i,j}$ overlapping user $i$ symbols in the duration of a single user $j$ symbol. The received signal is given by
\begin{equation}
    \mathbf{r}= \mathbf{H}_{i}\tildebf{s}_{i} + \mathbf{H}_{i}\mathbf{s}_j + \mathbf{w},
    \label{eq:6}
\end{equation}
and the $m$-th decoded user $i$ symbol, where $1 \leq m \leq \Delta_{i,j}$, is given by
\begin{equation}
    \mathbf{y}_{i,m} = \mathbf{F}_{i}\mathbf{R}_{\mathrm{cp},i}\mathbf{C}_m^{(i \leftarrow j)}\mathbf{r},
    \label{eq:7}
\end{equation}
where $\mathbf{C}^{(i \leftarrow j)} = \left[\mathbf{0}_{N_{T,i}\times(m-1)N_{T,i}}, \mathbf{I}_{N_{T,i}}, \mathbf{0}_{N_{T,i}\times(\Delta_{i,j}-m)N_{T,i}}\right]$ isolates the overlapping part of user $j$ symbol. Using (\ref{eq:1}) and (\ref{eq:6}), (\ref{eq:7}) can be expanded as
\begin{equation}
    \mathbf{y}_{i,m} =\mathbf{\Psi}_i\mathrm{diag}(\mathbf{x}_i)\mathrm{diag}(\mathbf{p}_i){\mathbf{d}_{i,m}}+ \mathbf{F}_{i}\mathbf{R}_{\mathrm{cp},i}\mathbf{C}_m^{(i \leftarrow j)}\mathbf{H}_{i}\mathbf{s}_j + \boldsymbol{\omega}_i.
    \label{eq:8}
\end{equation}
The interference matrix for symbol $m$ of user $i$ is therefore given by
\begin{equation}
        \boldsymbol{\Gamma}^{(i \leftarrow j)} = \mathbf{F}_{i}\mathbf{R}_{\mathrm{cp},i}\mathbf{C}_m^{(i \leftarrow j)}\mathbf{H}_{j}{\mathbf{A}}_{j}{{\mathbf{F}}_{j}^{\rm{H}}}\mathrm{diag}(\mathbf{x}_j)\mathrm{diag}(\mathbf{p}_j),
            \label{eq:9}
\end{equation}
which is an $N_i \times N_j$ matrix where the $(n,o)$-th element \({\Gamma}_{n,o}^{(i \leftarrow j)}\) defines the INI coefficient on subcarrier $n$ of user $i$ from the the $o$-th subcarrier of user $j$.

For both cases, the interference matrix can be used to calculate the mean-squared error (MSE) interference on the victim user
\begin{equation}
    \boldsymbol{{\gamma}}^{(i \leftarrow j)}(\mathbf{x}_j,\mathbf{p}_j) = \mathrm{diag}(\boldsymbol{\Gamma}^{(i \leftarrow j)}(\boldsymbol{\Gamma}^{(i \leftarrow j)})^{\mathrm{H}}),\label{eq:int}
\end{equation}
which is a vector of length \(N_i\) whose $j$-th element is equal to the MSE on the corresponding subcarrier of user $i$ due to interference from user $j$. The signal-to-interference-plus-noise ratio (SINR) on subcarrier $n$ of user $i$, expressed as a function of $\mathbf{x}$ and $\mathbf{p}$, is given by
\begin{equation}
    {\Lambda}_{i,n}(\mathbf{x},\mathbf{p})=\left(\frac{x_{i,n},p_{i,n}|h_{i,n}|^2}{\sum_{j=i+1}^{K}\gamma_n^{(i\leftarrow j)}(\mathbf{x}_j,\mathbf{p}_j)+{\sigma}^2}\right)\label{SINR},
\end{equation}
where $|h_{i,n}|^2$ is the channel gain on subcarrier $n$ of user $i$ and ${\sigma}^2$ denotes the noise power. The instantaneous achievable rate of user $i$ on subcarrier $n$ is given by
\begin{equation}
    R_{i,n}(\mathbf{x},\mathbf{p}) = \frac{B}{N_{i}}\log_2\left(1+{\Lambda}_{i,n}(\mathbf{x},\mathbf{p})\right),\label{eq:10}
\end{equation}
and the achievable sum-rate for the system can be expressed as
\begin{equation}
    R(\mathbf{x},\mathbf{p}) =  \sum_{i=1}^{K}\sum_{n = 1}^{N}R_{i,n}\label{eq:11}.
\end{equation}

\section{Resource Allocation}
 We define $\mathcal{N}=\{0,\dots,N_b-1\}$ as the set of subcarriers for the base numerology, i.e., the numerology with the smallest subcarrier spacing. The set of subcarriers for all other numerologies are subsets of $\mathcal{N}$ and are related by the relevant scaling factors. We then define the set of users that can use subcarrier $n \in \mathcal{N}$ as $\mathcal{K}_n=\{i:n \bmod q_i = 0\}.$ Our optimization problem aims to maximize the spectral efficiency of the system and can be formulated as
 \begin{subequations}
        \begin{alignat}{2}
        &\!\max_{\mathbf{x},\mathbf{p}}        &\qquad& \sum_{i=1}^{K}\sum_{n=0}^{N_{i}-1}R_{i,n}(\mathbf{x},\mathbf{p})\label{eq:optProb}\\
        &\text{subject to} &      & \sum_{n=0}^{N_{i}-1}p_{i,n}, \leq P_i\label{eq:constraint1}\\
        &                  &      & p_{i,n} \geq 0,\label{eq:constraint2}\\
        &                  &      & \sum_{n=0}^{N_{i}-1}R_{i,n} \geq R_{\rm{min}}, \forall i\in\{1,..,K\}, \label{eq:constraint3}\\
        &                  &      & x_{i,n} \in \{0,1\}, \label{eq:constraint4}\\
    &                  &      & \sum_{i\in \mathcal{K}_n}x_{i,\frac{n}{q_i}} \leq U, \forall_n \in \mathcal{N}, \label{eq:constraint5}
    \end{alignat}
\end{subequations}
where (\ref{eq:constraint1}) is the power constraint for each user, (\ref{eq:constraint2}) ensures the power on a subcarrier cannot be negative and (\ref{eq:constraint3}) ensures that each user achieves a rate $R_{\rm{min}}$. Constraints (\ref{eq:constraint4}) and (\ref{eq:constraint5}) ensure that the number of users allocated to any subcarrier does not exceed $U.$ 
\begin{algorithm}
    \caption{Subcarrier Allocation and Power Initialization}\label{SCA}
    \begin{algorithmic}[1]
    \State Initialize $\mathbf{X} = \mathbf{1}$
    \Repeat
    \State Use IWF for each user with SC allocation $\mathbf{X}$ to obtain power allocation $\mathbf{p}$
    \State Calculate $R(\mathbf{x},\mathbf{p})$ using (\ref{eq:10}) and (\ref{eq:11})
    \State For each $n \in \mathcal{N},$ $\mathcal{S}_n=\{i:p_{i,\frac{n}{q_i}}\geq 0$\}
    \State $U_{\mathrm{max}}=\max|\mathcal{S}_n|$ and $\mathcal{N}_{\mathrm{max}}=\{n:|\mathcal{S}_n|=U_{\mathrm{max}}$\}
    \If{$U_{\mathrm{max}} \geq U$} 
    \State $(i^*,n^*)= \arg\displaystyle\!\min_{n \in \mathcal{N}, i \in \mathcal{K}_n}(R^{-}_{i,n})$
    \State Set $x_{i^*,\frac{n^*}{q_{i^*}}}=0$
    \EndIf
    \Until{$U_{\mathrm{max}} = U$}
    \State \textbf{return} $\mathbf{X}^*=\mathbf{X}$ and $\mathbf{p}_{0} = \mathbf{p}$
    \end{algorithmic}
\end{algorithm}
It is clear that the objective function is non-convex due to the binary constraint (\ref{eq:constraint4}) and due to the nature of the interference term in (\ref{SINR}). This makes finding an optimal solution difficult. Instead we propose a two-stage solution. In Stage 1 we use an iterative greedy algorithm to allocate subcarriers to users and initialize the power allocation using iterative water filling (IWF). However, due to the presence of mixed numerologies, IWF does not provide an optimal power allocation. Therefore, in Stage 2, we use successive convex approximation to optimize the power allocation given the subcarrier allocation from Stage 1.

\subsection*{Stage 1: Subcarrier Allocation and Power Initialization}
The proposed greedy algorithm starts by relaxing constraint (\ref{eq:constraint5}) and setting $x_{i,n} = 1$ for all subcarriers of all users. Iterative water filling is is then used to allocate power for each user. We define the set of users with positive power on each subcarrier as $\mathcal{S}_n=\{i:p_{i,\frac{n}{q_i}}\geq 0$\}. This power allocation may lead to some subcarriers having too many active users, i.e., $|\mathcal{S}_n| > U.$ One user is then removed from the overloaded subcarriers in each iteration of the algorithm.
    
It is important to note that in MN-NOMA, users on one subcarrier can affect the rate of users on other subcarriers due to INI. The user to be removed from the overloaded subcarrier should therefore be the user which reduces the overall sum rate the least. This can be seen in Line 8 of Algorithm 1, where the user  chosen is the one which minimizes $R^{-}_{i,n} \triangleq R(\mathbf{x},\mathbf{p})-R(\mathbf{x}_{\setminus i,n},\mathbf{p})$ where $\mathbf{x}_{\setminus i,n}$ is simply $\mathbf{x}$ with $x_{i,n} = 0.$ The user which minimizes the reduction in overall rate is then de-allocated from subcarrier $n$ by setting $x_{i,n} = 0$. 
 
IWF is then performed again given this updated subcarrier allocation. This process is then repeated until there are no more overloaded subcarriers remaining. This provides the subcarrier allocation and initial power allocation for Stage 2. A detailed description of the algorithm can be found in Algorithm 1. 

\subsection*{Stage 2: Power Allocation Optimization}
 \begin{algorithm}
\caption{Power Allocation}\label{PA}
\begin{algorithmic}[1]
\State Initialize $t=0$; Initialize $\mathcal{M}_{i}$, $\boldsymbol{{\rho}}(t) = \boldsymbol{{\rho}}_0$ using Algorithm 1, $\mathbf{q}(t) = \log_2(\boldsymbol{{\rho}}_0)$, $\epsilon = 10^{-6}$
\Repeat
\State Update $\boldsymbol{{\gamma}}^{(i \leftarrow k)}$ using (\ref{eq:int}) for all users.
\State Update $\alpha_{i,n}$ and $\beta_{i,n}$ using (\ref{eq:bound}) for each subcarrier of each user
\State Solve (\ref{eq11}) to obtain $\mathbf{q}(t)$
\State $\boldsymbol{{\rho}}(t) \gets 2^{\mathbf{q}(t)}$
\State Use (\ref{eq:10}) and (\ref{eq:11}) to obtain $R(t)$
\State $t \gets t+1$
\Until{$R(t)-R(t-1)<\epsilon$}
\State \textbf{return} $\boldsymbol{{\rho}}^* = \boldsymbol{{\rho}}(t)$
\end{algorithmic}
\end{algorithm}
Given the subcarrier allocation $\mathbf{X}^*$ from Stage 1, we define the set of active subcarriers for each user as $\mathcal{ M}_{i} =\{n:x_{i,n}=1\}$ with elements $\{\mathcal{M}_{i,0}\ \mathcal{M}_{i,1}\ \dots\ \mathcal{M}_{i,N_{c,i}}\}$.  We then define the active power allocation vectors  ${\boldsymbol{{\rho}}_{i}} = \left[\sqrt{{p}_{i,\mathcal{M}_{i,0}}}\ \sqrt{{p}_{i,\mathcal{M}_{i,1}}}\ \dots\ \sqrt{{p}_{i,\mathcal{M}_{i,N_{c,i}}}} \right]^{\rm{T}}$ where $N_{c,i}$ is the number of active subcarriers for user $i$. The active power vector for the whole system is defined as $\boldsymbol{{\rho}} = \left[\boldsymbol{{\rho}}_{1}^{\rm{T}}\  \boldsymbol{{\rho}}_{2}^{\rm{T}}\ \dots\ \boldsymbol{{\rho}}_{K}^{\rm{T}} \right]^{\rm{T}}$.

We can now express the optimization problem as:
\begin{subequations}
    \begin{alignat}{2}
    &\!\max_{\boldsymbol{{\rho}}}        &\qquad& \sum_{i=1}^{K}\sum_{n \in \mathcal{ M}_{i}}R_{i,n}(\boldsymbol{{\rho}})\label{eq:optProb2}\\
    &\text{subject to} &      &\sum_{n \in \mathcal{ M}_{i}}p_{i,n} \leq P_i,\label{eq:constraint1.2}\\
    &                  &      &p_{i,n} \geq 0, \label{eq:constraint2.2}\\
    &                  &      &\sum_{n \in \mathcal{ M}_{i}}R_{i,n} \geq R_{\rm{min}}, \forall i\in\{1,..,K\}. \label{eq:constraint3.2}
    \end{alignat}
\end{subequations}
It is clear that the objective function is still non-convex due to (\ref{eq:10}), and in fact has a difference-of-convex (DC) structure. In addition, the change in the interference terms caused by the differing numerologies means that the method used for converting the single numerology NOMA problem in \cite{ZengUplinkNoma} into a convex problem does not work for MN-NOMA and an IWF approach no longer guarantees a global optimum. Instead we utilize the SCA approach \cite{Papand2006} to relax the DC structure of the problem. We utilize the lower bound
\begin{equation}
    \alpha\log(\Lambda)+\beta \leq \log(1+\Lambda), \label{eq:bound}
\end{equation}
which is tight for a given $\Lambda$ at the values of $\alpha=\frac {\Lambda }{1+\Lambda }$ and $\beta= \log (1+\Lambda) - \frac {\Lambda }{1+\Lambda }\log \Lambda$. We apply this bound to approximate (\ref{eq:optProb2}) as
\begin{equation}
    \sum_{i=1}^{K}\sum_{n \in \mathcal{ M}_{i}}\frac{B}{N_{i}}\left(\alpha_{i,n}\log_2({\Lambda}_{i,n}(\boldsymbol{{\rho}}))+\beta_{i,n}\right).\label{eq10}
\end{equation}
However, this is relaxed expression is still non-convex in $\boldsymbol{{\rho}}$. Therefore, we define vector $\mathbf{q}$ where $q_{i,j} = \log_2(p_{i,j})$ and use variable substitution to rewrite the objective function as
\begin{equation}
    \sum_{i=1}^{K}\!\!\sum_{n \in \mathcal{ M}_{i}}\!\!\!\!\bar{R}_{i,n}(\mathbf{q}) = \sum_{i=1}^{K}\!\!\sum_{n \in \mathcal{ M}_{i}}\!\!\!\frac{B}{N^{(i)}}\left(\alpha_{i,n}\log_2({\Lambda}_{i,n}(2^{\mathbf{q}}))+\beta_{i,n}\right),\label{eq10a}
\end{equation}
where $2^{\mathbf{q}}$ is an element-wise operation on the vector $\mathbf{q}$. The relaxed optimization problem can now be formulated as
\begin{subequations}
    \begin{alignat}{2}
    &\!\max_{\mathbf{q}}    &\qquad& \sum_{i=1}^{K}\sum_{n \in \mathcal{ M}_{i}}\bar{R}_{i,n}(\mathbf{q})\label{eq11}\\
    &\text{s.t.} &      & \sum_{n \in \mathcal{ M}_{i}}2^{q_{i,n}} \leq P_i,\label{eq:constraint1.3}\\
    &                  &      & \sum_{n \in \mathcal{ M}_{i}}\bar{R}_{i,n} \geq R_{\rm{min}}, \forall i\in\{1,..,K\}.\label{eq:constraint3.3}
    \end{alignat}
\end{subequations}
 Expanding the term $\log_2({\Lambda}_{i,n}({\boldsymbol{{\rho}}}))$ reveals it to be convex in $\mathbf{q}$. Additionally, the constraints are also convex. The relaxed optimization problem is therefore convex and can be solved using standard convex optimization tools such as CVX \cite{cvx}. The resulting power allocation procedure is summarized in Algorithm \ref{PA}.
\begin{figure}[t]
    \vspace{-5.5mm}
    \centering
    \includegraphics[width=\columnwidth]{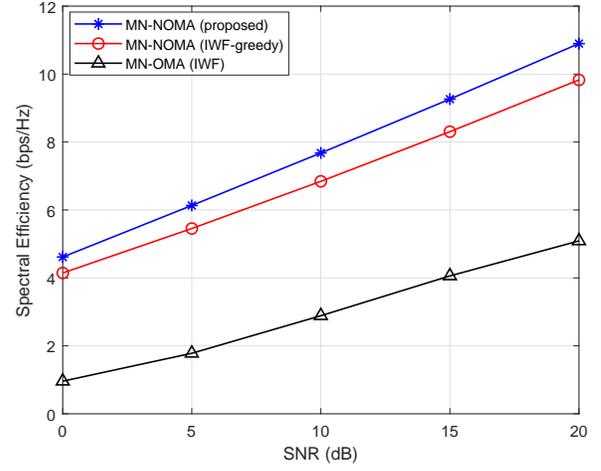}
    \caption{\label{fig:2} SE for MN-NOMA using our proposed algorithm, MN-NOMA using the IWF-Greedy algorithm from \cite{AlImariUpNOMA2015} and MN-OMA using IWF, at different SNR levels.}
\end{figure}
\section{Numerical Results}
This section presents numerical results to showcase the effectiveness of our proposed algorithm. As benchmarks, we consider the application of the IWF based greedy algorithm from \cite{AlImariUpNOMA2015} to the mixed numerology NOMA system, and we also consider an MN-OMA system using IWF for power allocation. For MN-OMA we assume that no guardbands exist between the users, which represents the best-case scenario for spectral efficiency. Each user has a power constraint of 1W per subcarrier to ensure that they use the same average power over a given time period. For the minimum rate constraint of the proposed algorithm, we set $R_{\rm{min}}=0.5$ bps/Hz. The extended vehicular A model was used to model the small-scale fading of each user, and Monte Carlo simulation is used to average the results over 1000 random channel instances. Three numerologies are used with DFT sizes of 512, 256 and 128, as per 5G NR specifications \cite{3gpp_TS38211}. The CP length is 7\% of the time domain symbol. For each instance in the Monte Carlo simulation, the number of users using each numerology is the same but the order in which they are decoded is random so that there is no numerology-based decoding order. A limit of $U=2$ users per subcarrier is adopted.

Fig. \ref{fig:2} shows the spectral efficiency of MN-NOMA using the proposed algorithm compared to the benchmark schemes for different SNR conditions for an 18-user case. It can be seen that the proposed algorithm outperforms the benchmark schemes in both cases, especially at high SNR. It can also be seen that both MN-NOMA schemes significantly outperform the MN-OMA scheme, confirming the benefits of MN-NOMA over MN-OMA.
\begin{figure}[t]
    \vspace{-5.5mm}
    \centering
    \includegraphics[width=\columnwidth]{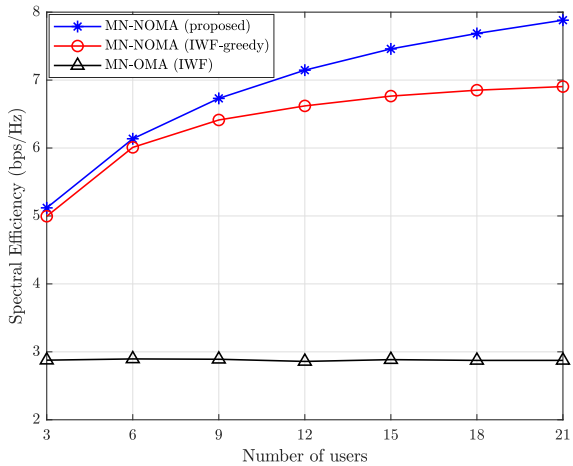}
    \caption{\label{fig:3} SE for MN-NOMA using our proposed algorithm, MN-NOMA using the IWF-Greedy algorithm from \cite{AlImariUpNOMA2015} and MN-OMA using IWF, at different different number of users.}
\end{figure}
 Fig. \ref{fig:3} shows the spectral efficiency of the proposed algorithm compared to the benchmark schemes with increasing number of users and a fixed SNR of 10dB. The proposed algorithm outperforms the benchmark schemes and this improvement becomes larger at a higher number of users.

\begin{figure}[t]
    \vspace{-5.5mm}
    \centering
    \includegraphics[width=\columnwidth]{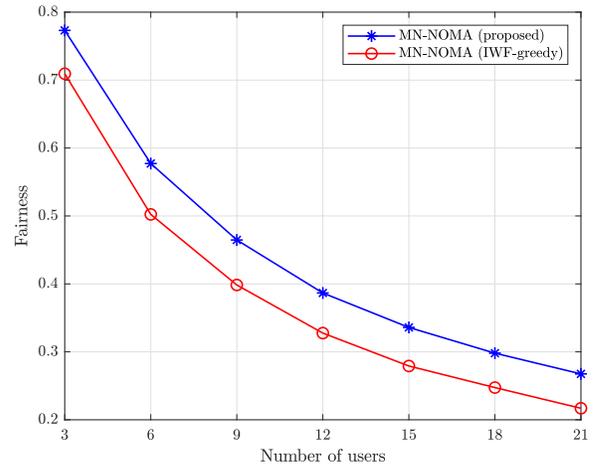}
    \caption{\label{fig:4} Fairness for MN-NOMA using our proposed algorithm and the IWF-Greedy algorithm from \cite{AlImariUpNOMA2015}, for different number of users.}
\end{figure}

 Fig. \ref{fig:4} compares the fairness of the proposed algorithm with that of the benchmark IWF scheme. Here the fairness is measured using Jain's fairness index
 \begin{equation}
     F = \frac{(\sum_{i=1}^{K}R_i)^2}{K\sum_{i=1}^{K}R_i^2},
 \end{equation}
and we consider a varying number of users for a fixed SNR of 10~dB. It can be observed from the figure that the proposed algorithm offers better fairness compared to the benchmark scheme due to the minimum rate constraint which ensures that each user is getting a fairer share of the sum rate.

\section{Conclusion}
In this paper we have studied the topic of resource allocation optimization for maximizing spectral efficiency in MN-NOMA. We have presented a generic system model for $K$-user uplink MN-NOMA and we have proposed a two-stage sub-optimal algorithm for maximizing spectral efficiency subject to a minimum rate constraint on each user. Stage 1 initializes power allocation and allocates users to subcarriers using a greedy algorithm and then Stage 2 then uses SCA to optimize power allocation. The superiority of the proposed method over benchmark schemes has been demonstrated numerically, in both spectral efficiency and fairness. The work presented in this paper can be further extended to consider optimal user decoding order and users with alternative waveforms.
\bibliographystyle{IEEEtran}
\bibliography{biblio}

\end{document}